\documentclass[english,aps,manuscript]{revtex4}
\usepackage[T1]{fontenc}
\usepackage[latin9]{inputenc}
\usepackage{graphicx}

\providecommand{\tabularnewline}{\\}

\usepackage{babel}

\begin{document}

\title{Pair Production of Tau Sneutrinos at Linear Colliders}

\author{V. Ar\i{}}

\email{vari@science.ankara.edu.tr}

\affiliation{Ankara University, Faculty of Sciences, Department of Physics, 06100,
Tandogan, Ankara}

\author{O. Çakir}

\email{ocakir@science.ankara.edu.tr}

\affiliation{Ankara University, Faculty of Sciences, Department of Physics, 06100,
Tandogan, Ankara }
\begin{abstract}
The pair production of tau sneutrinos in $e^{+}e^{-}$ collisions
and their subsequent decays are studied in a framework of the supersymmetric
extension of the standard model. We present an analysis for the parameter
space (BR vs. mass) which could be explored at the future high energy
$e^{+}e^{-}$ colliders.
\end{abstract}
\maketitle
The supersymmetry (SUSY) is considered to be a plausible candidate
for physics beyond the Standard Model (SM). The minimal supersymmetric
extension (MSSM) of the SM consists of the two-Higgs doublet extension
and the corresponding supersymmetric partners of the SM, whose spin
differs by one half unit. The SUSY and SM particles are distinguished
by a quantum number, $R$-parity, with $R=(-1)^{3B+L+2s}$ where $B$
and $L$ are the baryon and lepton numbers and $s$ is the spin. It
is commonly assumed that $R$-parity is conserved. As a consequence,
SUSY particles are produced in pairs, and the decays (cascade) into
SM particle plus SUSY particle ends with the lightest SUSY particle
(LSP).

The Large Hadron Collider (LHC) has a great potential to search for
supersymmetry, by virtue of its high beam energy and relatively large
sparticle production cross sections. In the hadron collisions, the
distribution of energy and momentum of the partons has a broad spectrum,
for the LSP searches only the missing transverse momentum (MPT) or
missing transverse energy (MET) become useful. Even some measurements
of sparticles could be possible at the Large Hadron Collider (LHC),
however, more precise determination \cite{Mizukoshi01,Freitas02}
of the underlying model parameters is necessary at future lepton colliders
operating with polarized beams \cite{Martin99,Moortgat08}. Furthermore,
being different from the hadron collisions, in the $e^{+}e^{-}$ collisions
the missing energy can be directly inferred from the center of mass
energy and the total energy of visible final state particles. In addition,
the leptonic beams have very small spread to correlate the missing
energy and momentum with the energy and momentum of the LSP's. The
spin-0 partners of the SM fermions (called sfermions) are the squarks,
sleptons and sneutrinos. The sleptons and squarks of the third family
are particularly interesting since their phenomenology is different
from that of other two families. This is expected due to the mixing
between the left-handed and right-handed components, and relatively
their large Yukawa couplings. In future experiments, study of sneutrinos
\cite{Bartl98,Choudhury05} and their mixings might also be interesting
\cite{Grossman97,Obara06,Oshimo04}.

The sneutrino mass limit is given from the direct searches $m_{\tilde{\nu}}>94$
GeV \cite{Amsler08}, with the assumption of mass degeneracy and the
presence of only the left handed sneutrinos $\tilde{\nu}_{L}$. From
the results obtained by the LEP Collaborations on the invisible width
of the $Z$ boson ($\Delta\Gamma_{inv.}<2.0$ MeV) the limit on the
sneutrino mass is given as $m_{\tilde{\nu}}>44.7$ GeV.

In this work, we study the pair production of tau sneutrinos in $e^{+}e^{-}$
collisions, as well as their subsequent decays into tau lepton plus
lighter chargino, and cascade ending with neutralinos (assuming LSP).
We present an analysis for the parameter space (BR vs. mass) which
could be explored at the future high energy $e^{+}e^{-}$ collider,
namely the Compact Linear Collider (CLIC) in two beam acceleration
technology allowing the preferable center of mass energy 3 TeV with
$L\approx10^{35}$ cm$^{-2}$s$^{-1}$ \cite{Assmann00}.

The coupling of the tau sneutrinos with the $Z$ boson is expressed
by the interaction

\begin{equation}
L=-\frac{i\sqrt{g^{2}+g'^{2}}}{2}(\tilde{\nu}_{\tau}^{*}\partial^{\mu}\tilde{\nu}_{\tau}-\partial^{\mu}\tilde{\nu}_{\tau}^{*}\tilde{\nu}_{\tau})Z_{\mu}\label{eq:1}\end{equation}
where $g$ and $g'$ are the gauge coupling constants. For the sneutrino
pair production the cross section is given by 

\begin{equation}
\sigma\left(e^{+}e^{-}\rightarrow\tilde{\nu_{\tau}}\bar{\tilde{\nu_{\tau}}}\right)=\frac{\alpha^{2}\pi}{192c_{w}^{4}s_{w}^{4}}\frac{s\left(\vartheta_{e}^{2}+a_{e}^{2}\right)}{\left(s-m_{z}^{2}\right)^{2}+m_{z}^{2}\Gamma_{z}^{2}}(1-\frac{4m_{\tilde{\nu}}^{2}}{s})^{3/2}\label{eq:2}\end{equation}
where $\vartheta_{e}=-1+4s_{w}^{2}$ and $a_{e}=-1$. The sneutrino
mass is given by $m_{\tilde{\nu}}^{2}=m_{\tilde{L}}^{2}-(m_{Z}^{2}/2)\cos2\beta$.
The tau sneutrino decays mainly to a chargino and a tau lepton. The
interaction term for the tau sneutrino-chargino-tau vertex is given
by

\begin{equation}
L=-g\bar{\tau}(C_{i\tau}^{L}P_{L}+C_{i\tau}^{R}P_{R})\tilde{\chi}_{i}^{-}\tilde{\nu}_{\tau}+h.c.\label{eq:3}\end{equation}
where $C_{i\tau}^{L}=V_{i1}^{*}$ and $C_{i\tau}^{R}=-y_{\tau}U_{i2}$
and the Yukawa coupling $y_{\tau}=m_{\tau}/\sqrt{2}m_{W}\cos\beta$.
We use the matrices $\mathbf{U}$ and $\mathbf{V}$ to diagonalize
chargino mass matrix. The chargino-neutralino-$W$ boson interaction
can be written as 

\begin{equation}
L=gW_{\mu}^{-}\bar{\tilde{\chi}}_{i}^{0}\gamma^{\mu}(D_{ik}^{L}P_{L}+D_{ik}^{R}P_{R})\tilde{\chi}_{k}^{+}+h.c.\label{eq:4}\end{equation}
where couplings $D_{ik}^{L}=-Z_{i4}V_{k2}^{*}/\sqrt{2}+Z_{i2}V_{k1}^{*}$
and $D_{ik}^{R}=Z_{i3}^{*}U_{k2}/\sqrt{2}+Z_{i2}^{*}U_{k1}$, the
interactions and notations can be found in \cite{Drees05}. Here,
we use the $4\times4$ matrices $\mathbf{Z}$ which diagonalize neutralino
mass matrix. 

We study the process $e^{+}e^{-}\to\tilde{\nu}_{\tau}\bar{\tilde{\nu}_{\tau}}$
with subsequent SUSY decays $\tilde{\nu}_{\tau}\to\tau^{-}\tilde{\chi_{1}}^{+}$and
$\tilde{\chi}^{+}\to W^{+}\tilde{\chi}_{1}^{0}$. Benchmark points
$\alpha,\beta,\gamma,\delta$ and the mass spectra (in GeV) for third
family sfermions, lightest chargino and neutralino, as calculated
using SuSpect \cite{Djouadi02} are given in Table \ref{tab:1}. Here,
the scenarios $\alpha,\beta,\gamma$ belongs to non-universal Higgs
mass (NUHM) models, while $\delta$ belongs to gravitino dark matter
(GDM). The point $\alpha$ has parameters $m_{0}=210,$ $m_{1/2}=285$,
$\tan\beta=10$; point $\beta$ has $m_{0}=230$, $m_{1/2}=360$,
$\tan\beta=10$; point $\gamma$ has $m_{0}=330$, $m_{1/2}=240$,
$\tan\beta=20$; and point $\delta$ has $m_{0}=500$, $m_{1/2}=750$,
$\tan\beta=10$. All these points have also the parameters sign$\mu=+1$
and $A_{0}=0$. At the points $\alpha$, $\beta$, $\gamma$ and $\delta$,
tau sneutrino has mass values $274.9$ GeV, $324.6$ GeV, $353.9$
GeV and $695.2$ GeV, respectively. The branching ratios for tau sneutrino
into neutralino+tau and chargino+tau neutrino are given in Table \ref{tab:2}
for a set of benchmark supersymmetric scenarios \cite{Roeck07}. The
tau sneutrino mostly decays to chargino and tau lepton for the $\alpha$
and $\gamma$ benchmark points. While it is comparable to the decay
into neutralino for the points $\beta$ and $\delta$. The reaction
is then observed as $e^{+}e^{-}\to\tau^{+}\tau^{-}W^{-}W^{+}+\not\! E_{T}$
where $\tau$-leptons has an energy spectrum with two edges. An analysis
on the sneutrino cascade decays as probe of the chargino spin properties
and CP violation was studied in \cite{Aguilar05}. Above the threshold
center of mass energy the tau sneutrinos are producedIn the $e^{+}e^{-}$
collisions. A threshold scan for the energy could help to identify
the model parameters. Fig. \ref{fig:fig3} shows the cross sections
for tau sneutrino pair production depending on the center of mass
energies covering relevant threshold regions. For the points $\alpha$,
$\beta$, $\gamma$ and $\delta$ the cross sections show peaks at
the center of mass energies $900$ GeV, $1050$ GeV, $1150$ GeV and
$2200$ GeV, respectively. At the high energies, preferable energy
of the CLIC ($\sqrt{s}=3$ TeV), the cross sections for different
benchmark points tend to converge around $1$ fb, which will yield
100 events at a luminosity of $L_{int}=10^{5}$ pb$^{-1}$.

\begin{table}[t]
\caption{Benchmark points and the mass spectra (in GeV) for third family sfermions,
lightest chargino and neutralino, as calculated using SuSpect \cite{Djouadi02}.
\label{tab:1}}

\begin{tabular}{ccccc}
\hline 
Model & ~~ ~~~$\alpha$~~ ~~~ & ~ ~~~~$\beta$~ ~~~~ & ~ ~~~~$\gamma$ ~~~~~ & ~ ~~~~$\delta$ ~~~~ ~\tabularnewline
\hline
$m_{1/2}$ & 285 & 360 & 240 & 750\tabularnewline
$m_{0}$ & 210 & 230 & 330 & 500\tabularnewline
$tan\beta$ & 10 & 10 & 20 & 10\tabularnewline
$sign(\mu)$ & + & + & + & +\tabularnewline
$A_{0}$ & 0 & 0 & 0 & 0\tabularnewline
$m_{t}$ & 175.0 & 175.0 & 175.0 & 175.0\tabularnewline
\hline 
Masses &  &  &  & \tabularnewline
\hline
$\tilde{\chi}_{1}^{0}$ & 112.4 & 144.7 & 94.2 & 315.8\tabularnewline
$\tilde{\chi}_{1}^{\pm}$ & 208.9 & 271.5 & 174.4 & 596.6\tabularnewline
$\tilde{\tau}_{1}$ & 232.2 & 262.4 & 323.0 & 564.0\tabularnewline
$\tilde{\tau}_{2}$ & 288.9 & 336.4 & 369.6 & 700.6\tabularnewline
$\tilde{\nu}_{\tau}$ & 274.9 & 324.6 & 353.9 & 695.2\tabularnewline
$\tilde{t}_{1}$ & 481.2 & 599.2 & 443.4 & 1214.0\tabularnewline
$\tilde{t}_{2}$ & 659.6 & 789.1 & 607.2 & 1484.0\tabularnewline
$\tilde{b}_{1}$ & 602.5 & 739.3 & 551.6 & 1459.0\tabularnewline
$\tilde{b}_{2}$ & 637.1 & 777.2 & 600.4 & 1525.0\tabularnewline
\hline
\end{tabular}
\end{table}

\begin{table}[t]
\caption{Branching ratios for tau sneutrino at different supersymmetry parameter
space. \label{tab:2}}

\begin{tabular}{|c|c|c|c|c|}
\hline 
 & $\alpha$ & $\beta$ & $\gamma$ & $\delta$\tabularnewline
\hline
\hline 
BR$\left(\tilde{\nu_{\tau}}\rightarrow\nu_{\tau}\tilde{\chi_{1}}^{0}\right)${[}\%] & 31.5 & 41.5 & 16.1 & 39.1\tabularnewline
\hline 
BR$\left(\tilde{\nu_{\tau}}\rightarrow\nu_{\tau}\tilde{\chi_{2}}^{0}\right)${[}\%] & 20.5 & 18.2 & 24.4 & 18.6\tabularnewline
\hline 
BR$\left(\tilde{\nu_{\tau}}\rightarrow\tau^{-}\tilde{\chi_{1}}^{+}\right)${[}\%] & 48.0 & 40.3 & 59.6 & 37.9\tabularnewline
\hline
\end{tabular}
\end{table}

\begin{figure}
\includegraphics{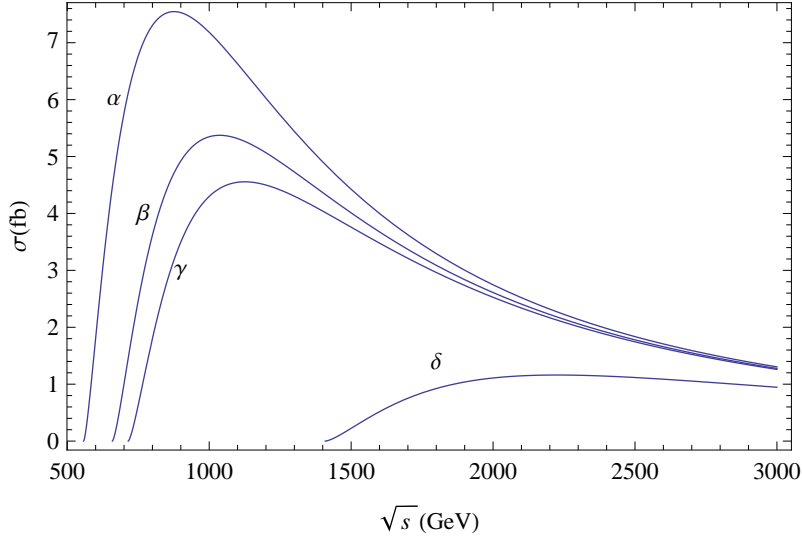}

\caption{Cross sections for pair production of tau sneutrinos depending on
the center of mass energy for different SUGRA points. \label{fig:fig1}}

\end{figure}

\begin{figure}
\includegraphics[scale=0.3]{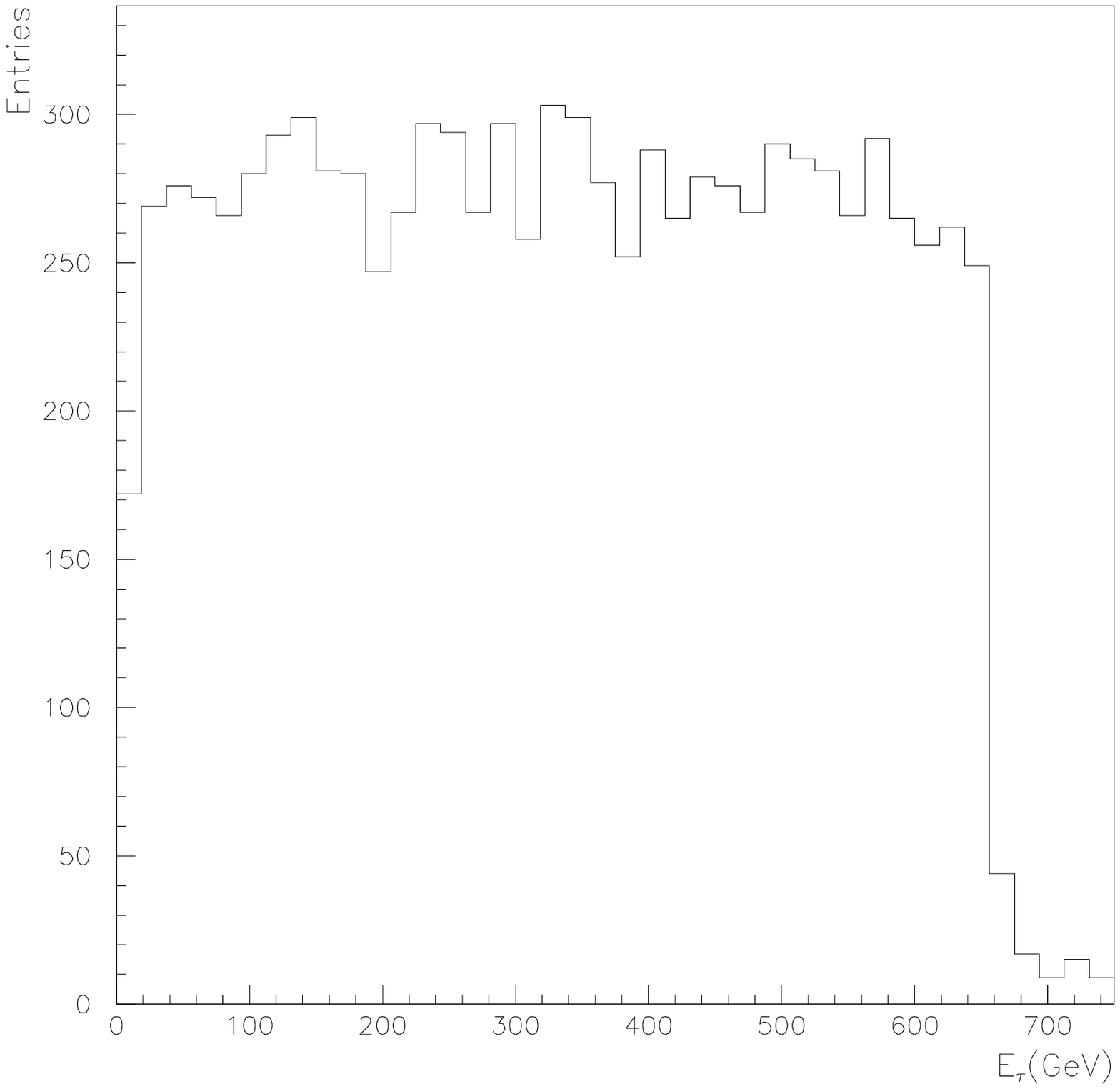}\includegraphics[scale=0.3]{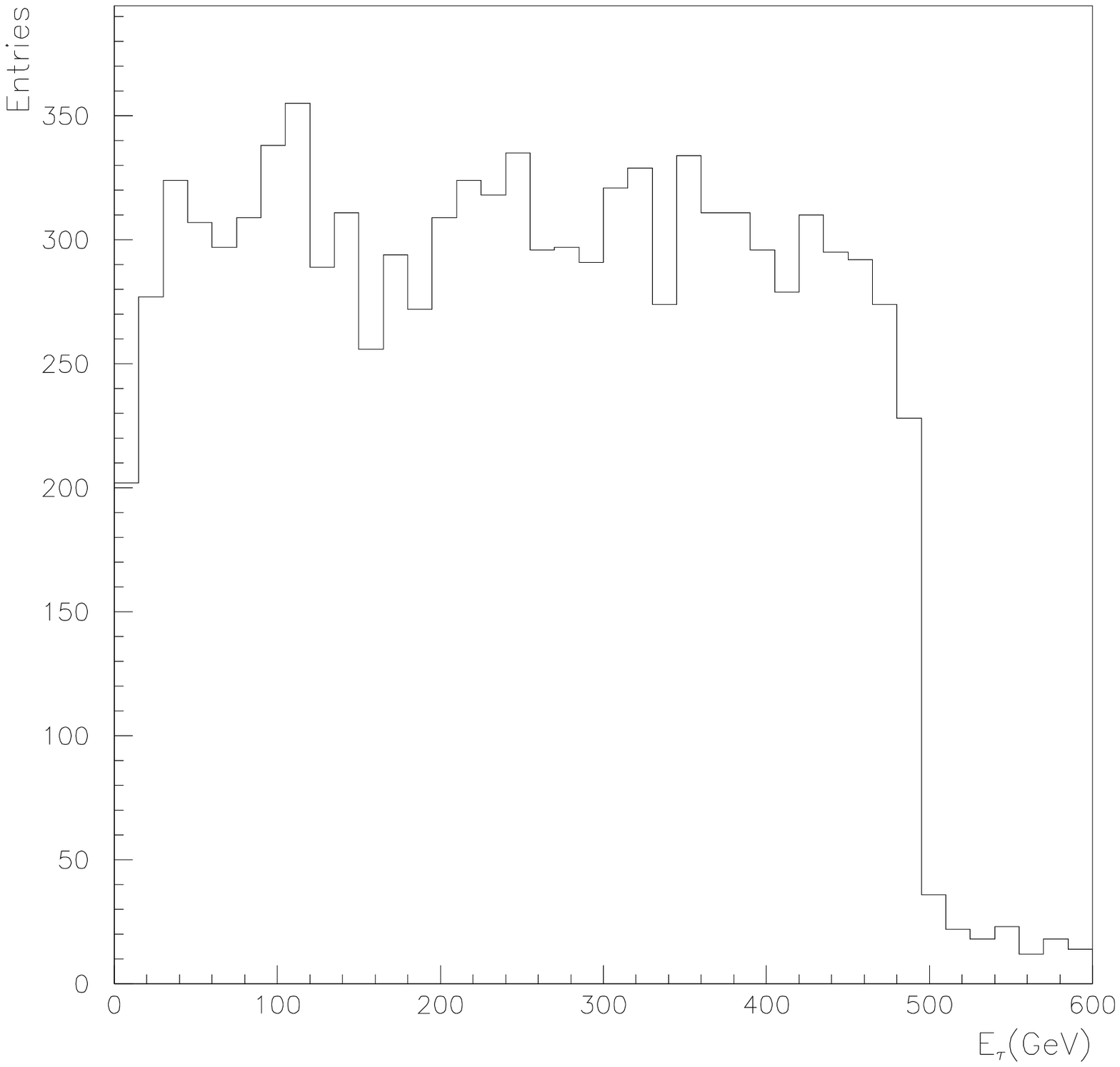}

\includegraphics[scale=0.3]{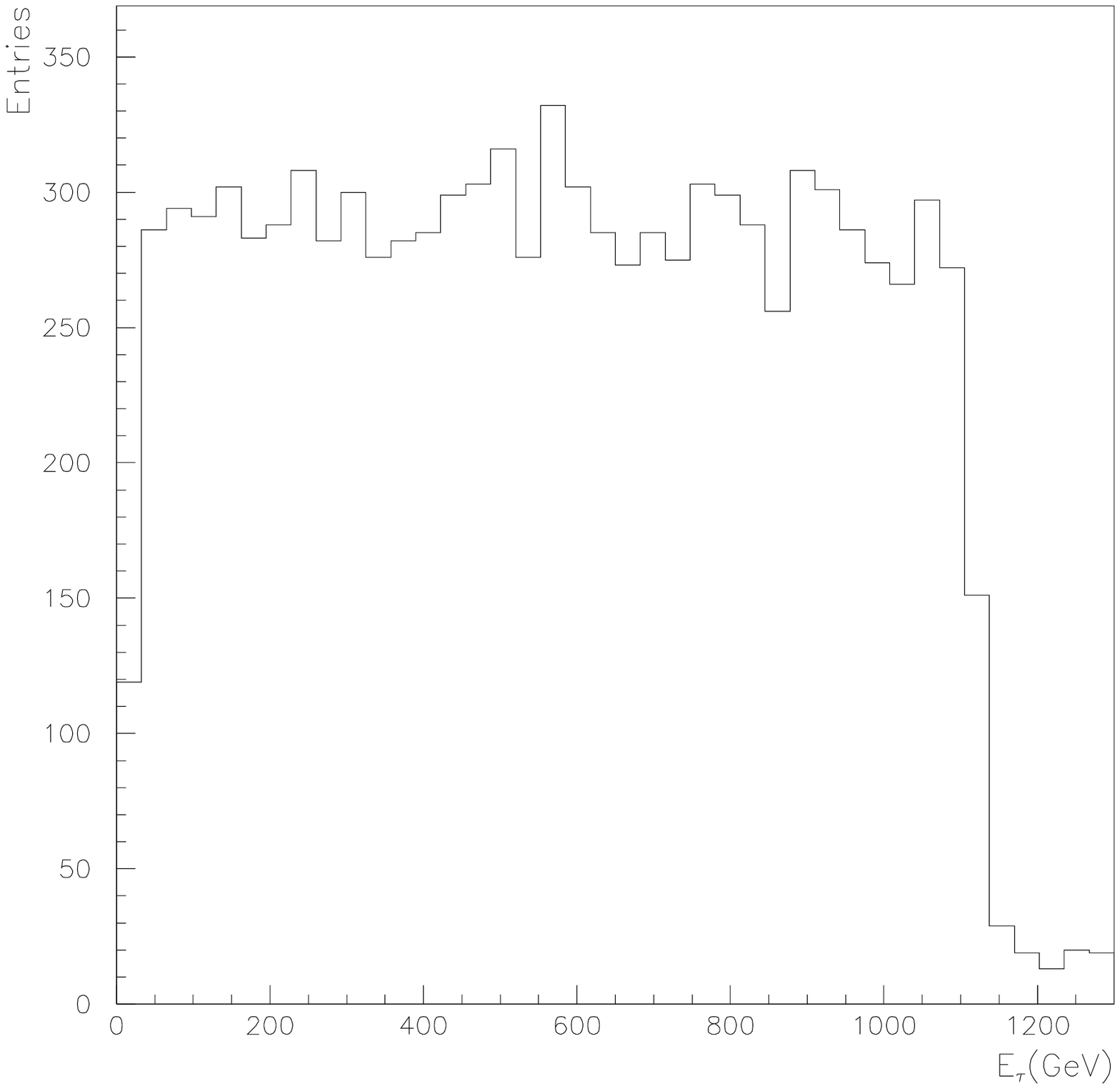}\includegraphics[scale=0.3]{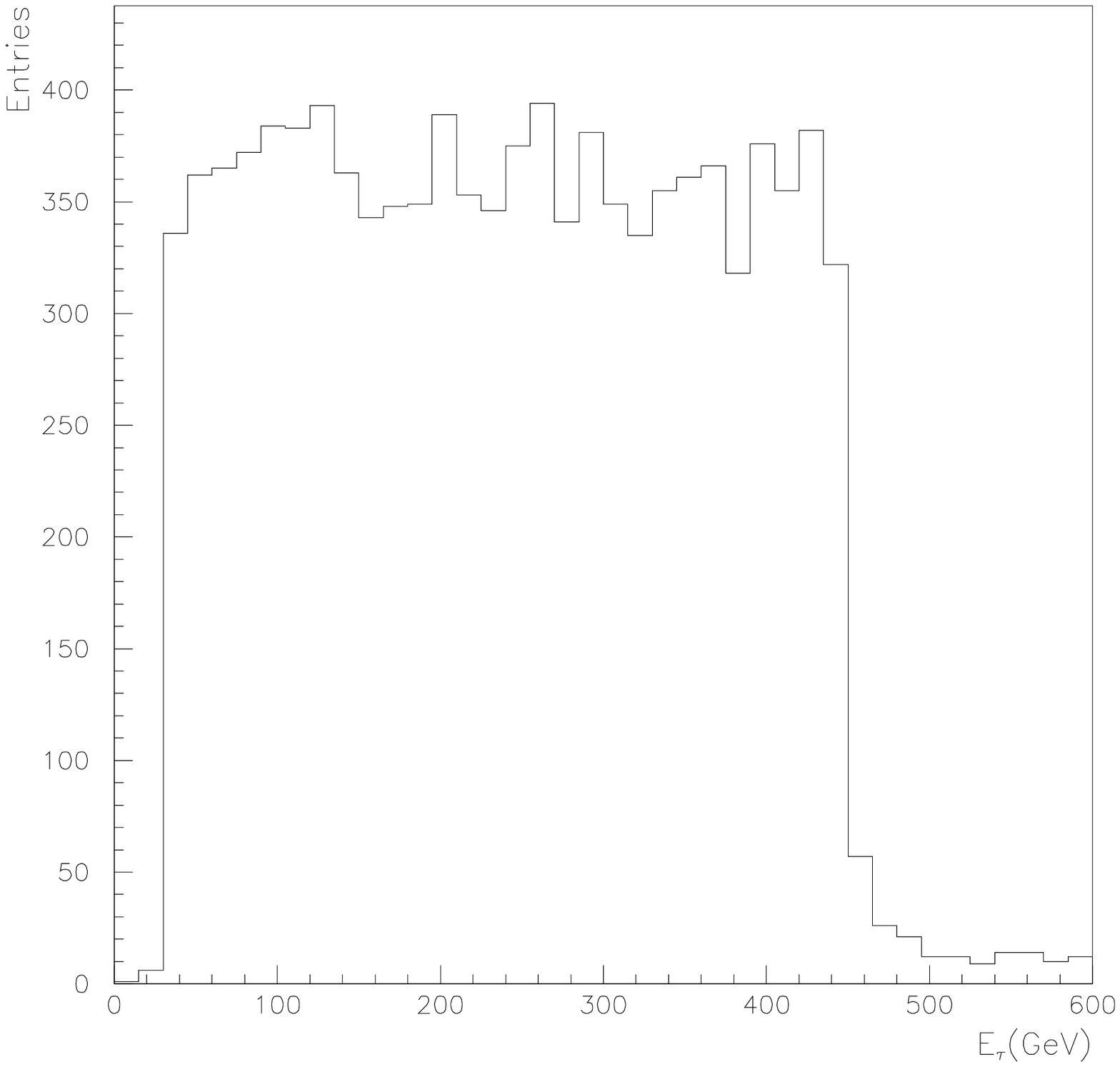}

\caption{The energy distributions of final state tau lepton for different SUGRA
points $\alpha$, $\beta$, $\gamma$ and $\delta$, from left/up
to right/down, respectively. \label{fig:fig2}}

\end{figure}

\begin{figure}
\includegraphics{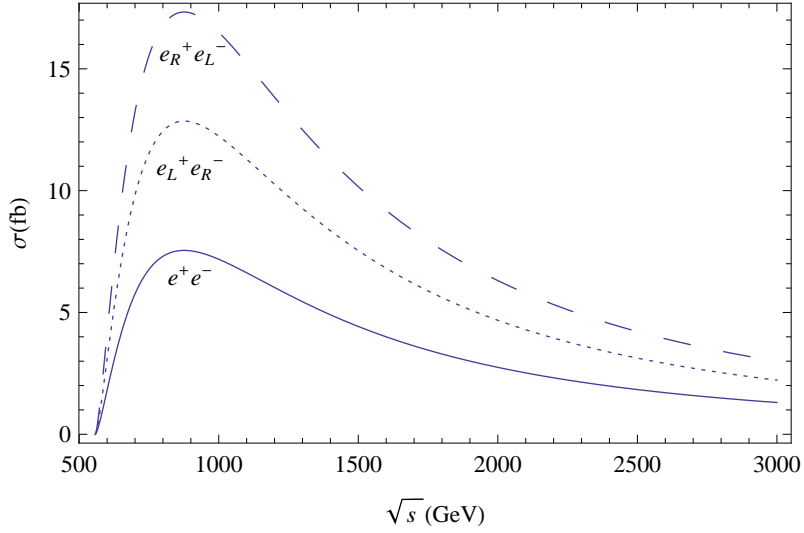}\caption{The cross section depending on the center of mass energy range for
point $\alpha$. The lower curve denotes unpolarized case, while upper
shows $RL$ and $LR$ polarization for positron and electron beams.\label{fig:fig3}}

\end{figure}

Since the tau sneutrino has spin-0 it decays isotropically, and the
boost of an isotropic distribution become a flat distribution in energy.
Therefore, we expect the tau energy spectrum nearly flat between some
kinematic endpoints \cite{Peskin08}. The two endpoints energies $E_{\pm}$
of the spectrum are related to the tau sneutrino and chargino masses
and the $\tilde{\nu}_{\tau}$ boost as

\begin{equation}
E_{\pm}=\frac{\sqrt{s}}{2m_{\tilde{\nu}}}(1\pm\sqrt{1-4m_{\tilde{\nu}}^{2}/s})\frac{m_{\tilde{\nu}}^{2}-m_{\tilde{\chi_{1}}}^{2}}{2m_{\tilde{\nu}}}\label{eq:5}\end{equation}
where $\sqrt{s}$ is the collision center of mass energy for the process
$e^{+}e^{-}\to\bar{\tilde{\nu}}_{\tau}\tilde{\nu}_{\tau}\to\tau^{+}\tilde{\chi}_{1}^{-}\tau^{-}\tilde{\chi}_{1}^{+}$.
This process provide a source of about fully polarized charginos,
$\tilde{\chi}_{1}^{+}$ with negative helicity, and $\tilde{\chi}_{1}^{-}$
with positive helicity. Using the benchmark points we calculate the
endpoint energies as: $E_{-}(E_{+})=6.3(656.9)$ for $\alpha$; $E_{-}(E_{+})=6.6(492.5)$
for $\beta$; $E_{-}(E_{+})=17.4(1121.4)$ for $\gamma$; $E_{-}(E_{+})=30.8(449.6)$
for $\delta$. The PYTHIA \cite{Sjostrand06} is used for event generatation
and the energy distributions of the final state tau lepton, which
are shown in Fig. \ref{fig:fig2} for different SUGRA points ($\alpha,\beta,\gamma,\delta$).
We may determine the tau sneutrino mass by using the ratio of the
two edge energies. If only the high energy edge is available, two-parameter
fit can be used to obtain the required accuracy in the mass measurement.

The decay width of tau sneutrino is approximately determined by two-body
problem as defined by

\begin{eqnarray}
\Gamma(\tilde{\nu}_{\tau} & \to & \tau^{-}\tilde{\chi}_{1}^{+})=\frac{g^{2}}{16\pi}m_{\tilde{\nu_{\tau}}}\sqrt{(1-\frac{(m_{\tau}+m_{\tilde{\chi}_{1}})^{2}}{m_{\tilde{\nu}}^{2}})(1-\frac{(m_{\tau}-m_{\tilde{\chi_{1}}})^{2}}{m_{\tilde{\nu}}^{2}})}\nonumber \\
 &  & \times\left[(|C_{L}|^{2}+|C_{R}|^{2}\frac{m_{\tau}^{2}}{2\cos^{2}\beta m_{W}^{2}})(1-\frac{m_{\tau}^{2}+m_{\tilde{\chi_{1}}}^{2}}{m_{\tilde{\nu}}^{2}})-2\sqrt{2}C_{L}C_{R}^{*}\frac{m_{\tilde{\chi_{1}}}m_{\tau}^{2}}{\cos\beta m_{W}m_{\tilde{\nu}}^{2}}\right]\label{eq:6}\end{eqnarray}

The cross section for the process $e^{+}e^{-}\to\tilde{\nu}_{\tau}\bar{\tilde{\nu}}_{\tau}$
has a special characteristics of scalar production proportional to
$\beta^{3}$. For the pair production of tau sneutrinos, the required
initial state has angular momentum $J_{z}=1$ state, since the final
state particles have spin-0, they must be produced in a $P$-wave,
and the production cross section increases as $\beta^{3}$ near threshold
(see Fig. \ref{fig:fig1}), which is in contrast to fermion pair production,
where the cross section increases as $\beta$. The measurement of
this behaviour shows that sneutrinos are scalars, and may allow precise
mass measurements at the $0.1$ GeV level assuming at least a luminosity
of $100$ fb$^{-1}$. The sneutrino masses can be determined from
a threshold scan. The unpolarized cross sections for the signal and
relevant background processes are given in Table \ref{tab:3}. Here,
the cross section for the signal shows the resulting values after
sneutrino and chargino decays ending the chain to neutralinos (LSP).
While the background have the cross sections as given by the process,
for further decays these values can be multiplied with the corresponding
branching ratios to compare with the signal.

\begin{table}
\caption{The unpolarized cross sections for signal and relevant background
at $\sqrt{s}$=3 TeV. \label{tab:3} }

\begin{tabular}{|c|c|c|c|c|c|c|c|c|}
\hline 
 & \multicolumn{4}{c|}{Signal} & \multicolumn{3}{c|}{Background} & \tabularnewline
\hline
\hline 
Benchmarks & $\alpha$ & $\beta$ & $\gamma$ & $\delta$ & $W^{+}W^{-}\tau^{+}\tau^{-}$ & $W^{+}W^{-}W^{+}W^{-}$ & $W^{+}W^{-}ZZ$ & $W^{+}W^{-}Z$\tabularnewline
\hline 
$\sigma$(fb) & 0.35 & 0.23 & 0.52 & 0.14 & 0.77 & 1.45 & 1.16 & 32.6\tabularnewline
\hline
\end{tabular}
\end{table}

\begin{figure}
\includegraphics{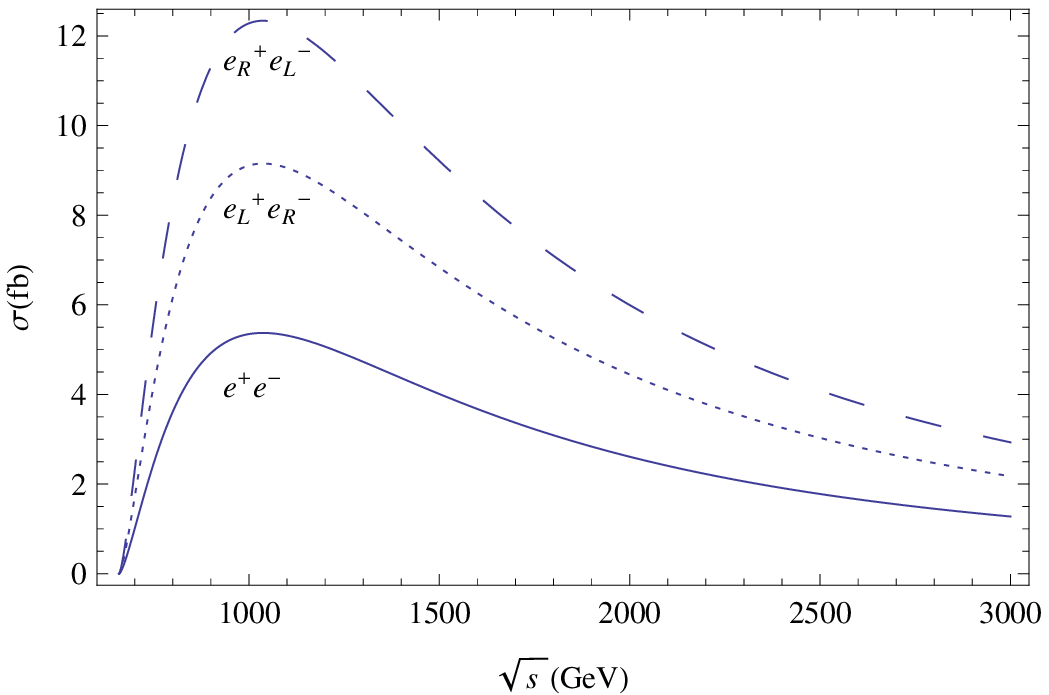}\caption{The same as Fig. \ref{fig:fig3}, but for point $\beta$. \label{fig:fig4}}

\end{figure}

\begin{figure}
\includegraphics{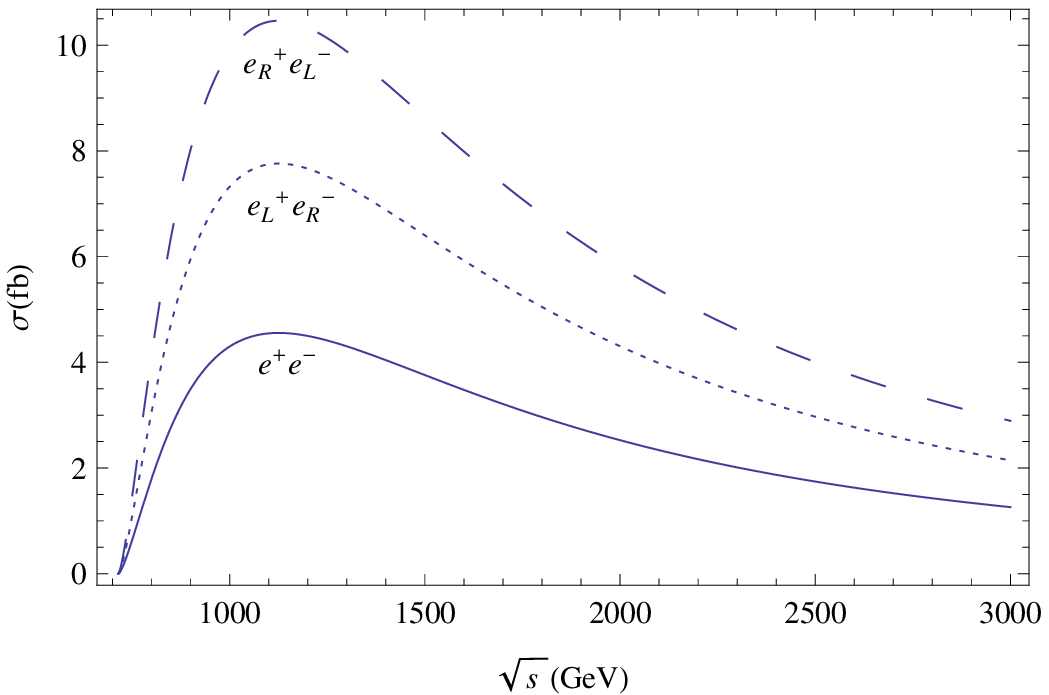}\caption{The same as Fig. \ref{fig:fig4}, but for point $\gamma$. \label{fig:fig5}}

\end{figure}

\begin{figure}
\includegraphics{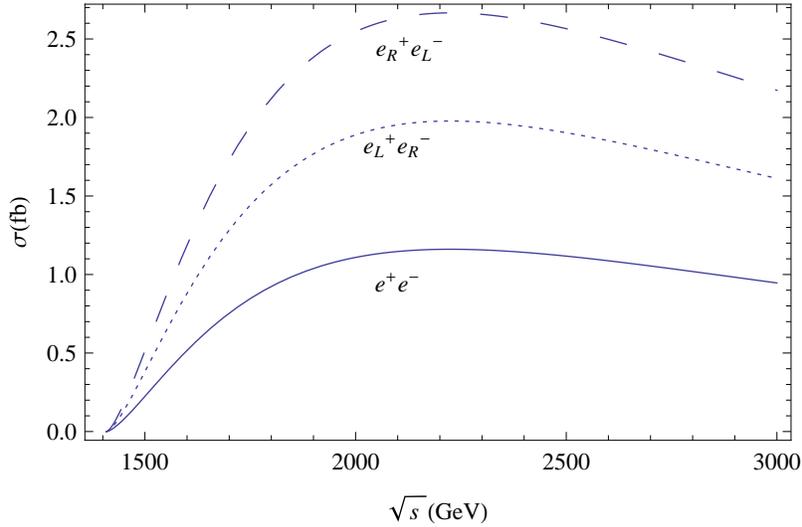}\caption{The same as Fig. \ref{fig:fig5}, but for point $\delta$. \label{fig:fig6}}

\end{figure}

The total cross sections for pair production of tau sneutrinos at
the energy range of $\sqrt{s}=500-3000$ GeV are shown in Fig. \ref{fig:fig3},
\ref{fig:fig4}, \ref{fig:fig5}, \ref{fig:fig6} for the points $\alpha$,
$\beta$, $\gamma$ and $\delta$. For point $\alpha$, the maximum
of the cross section is about $7$ fb for unpolarized beams at $\sqrt{s}\approx900$
GeV. We obtain larger cross section $16$ fb at maximum for the $e_{R}^{+}e_{L}^{-}$
polarization. At the other points $\beta$ and $\gamma$, the cross
section show maximum around $\sqrt{s}\approx1000$ GeV, but they are
different in magnitude. Since the $\delta$ point has larger gaugino
and scalar mass values, the cross section has maximum at $\sqrt{s}\approx2300$
GeV. The machine operating at $\sqrt{s}=3000$ GeV will produce tau
sneutrino with a cross section $0.9$ fb for unpolarized beams.

\begin{figure}
\includegraphics[scale=0.25]{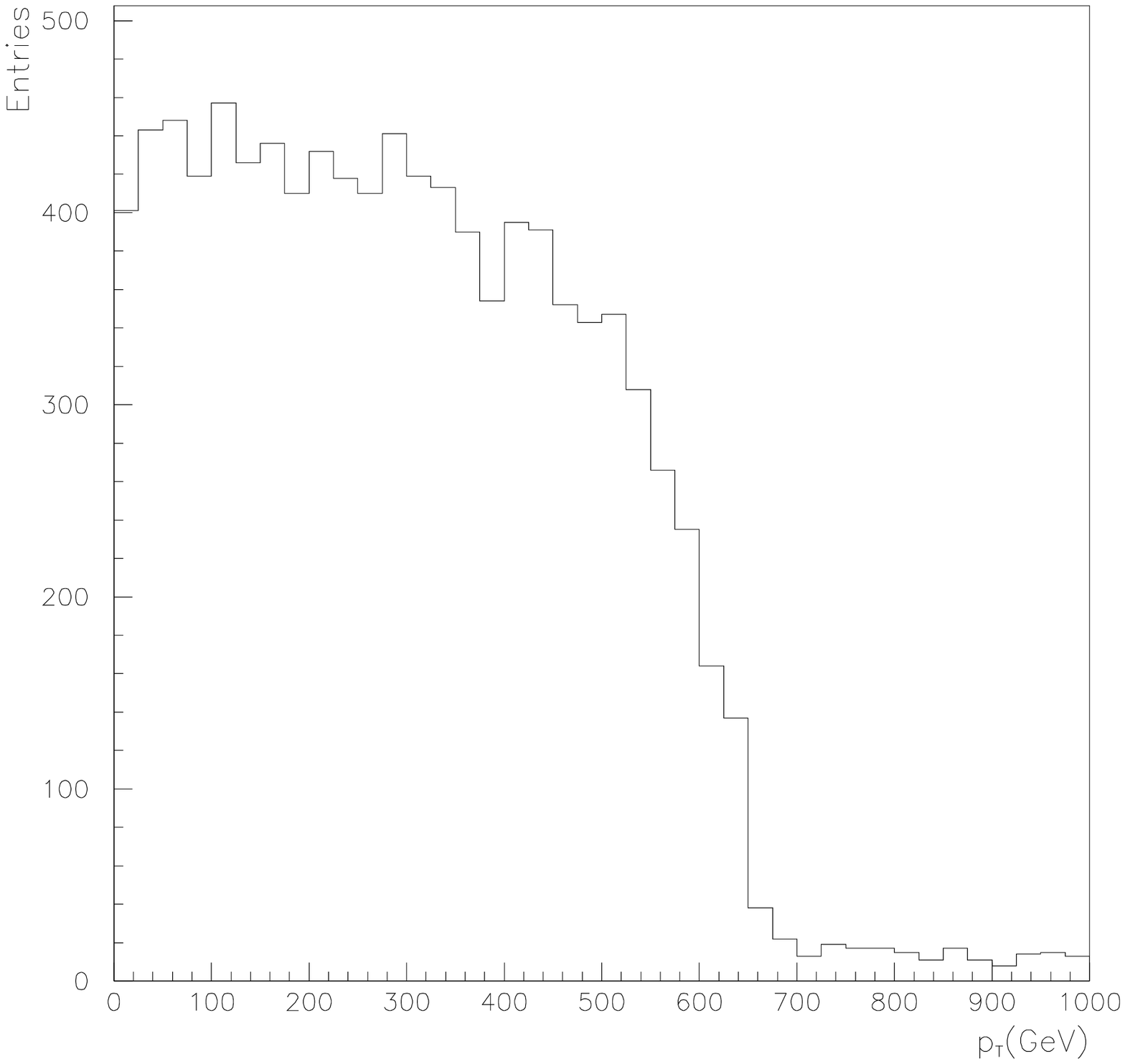}\includegraphics[scale=0.25]{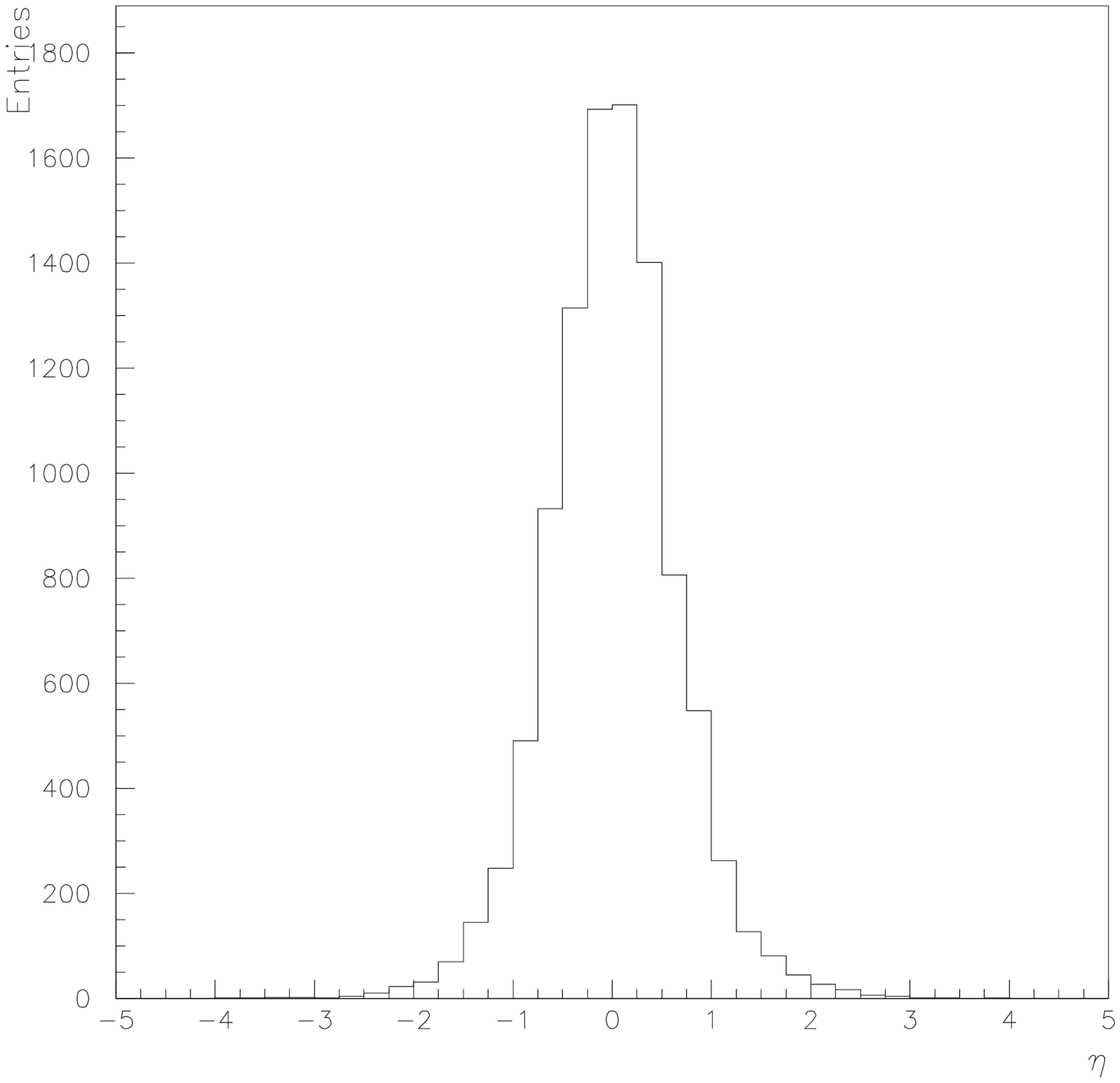}\includegraphics[scale=0.25]{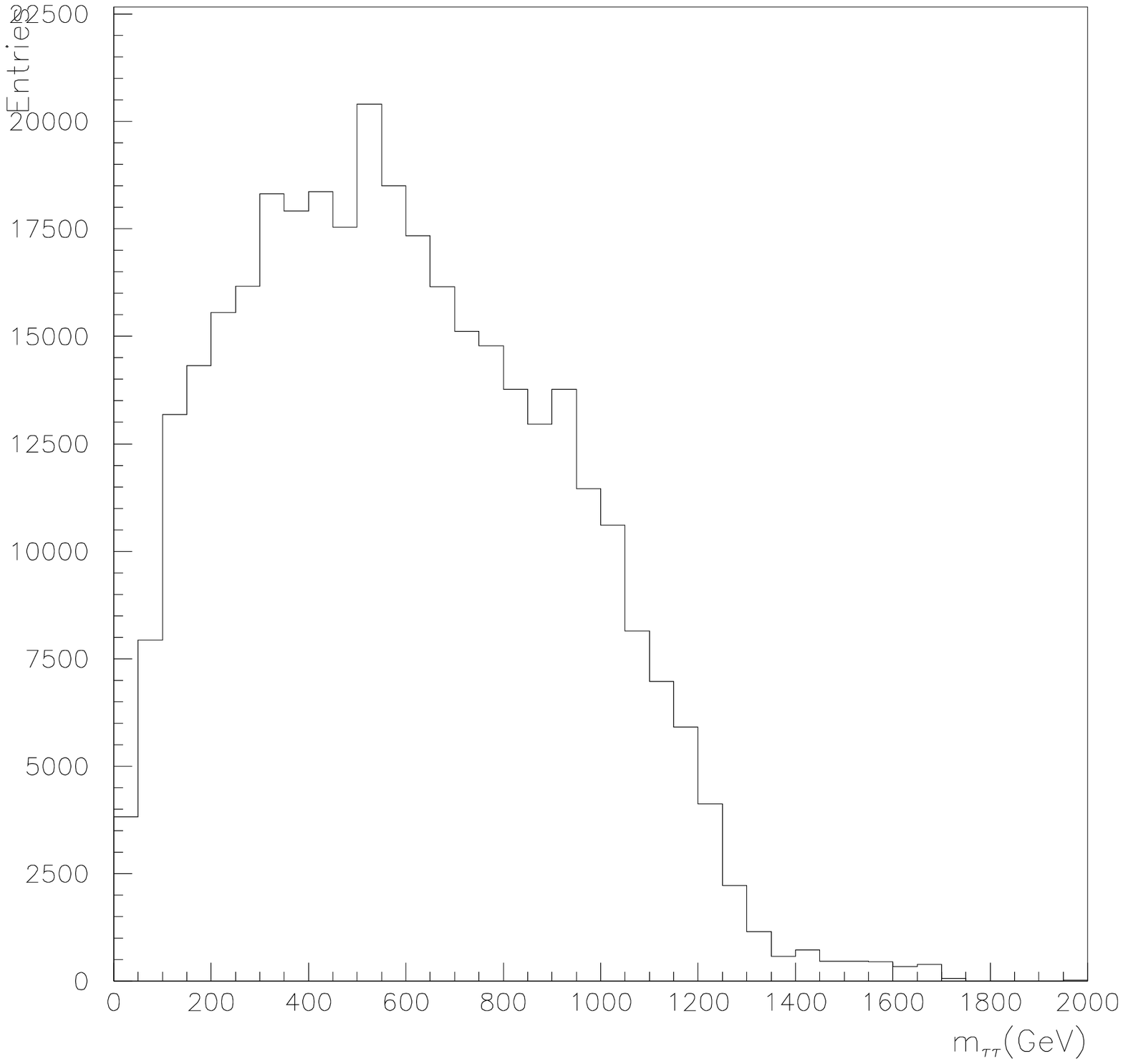}

\caption{The $p_{T}$ distribution, $\eta$ distribution and $m_{\tau\tau}$
distribution for the point $\alpha$. \label{fig:fig7}}

\end{figure}

\begin{figure}
\includegraphics[scale=0.25]{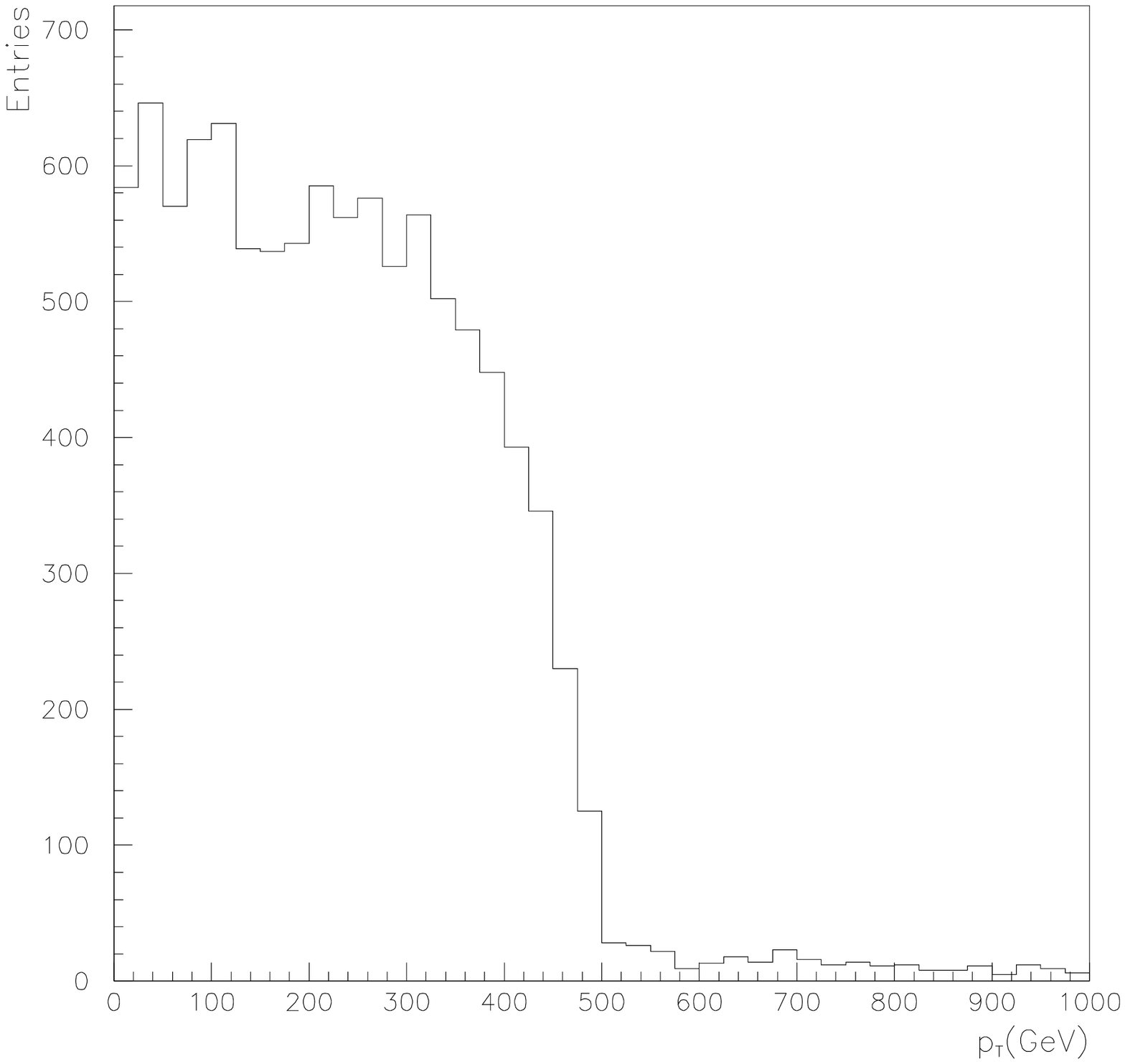}\includegraphics[scale=0.25]{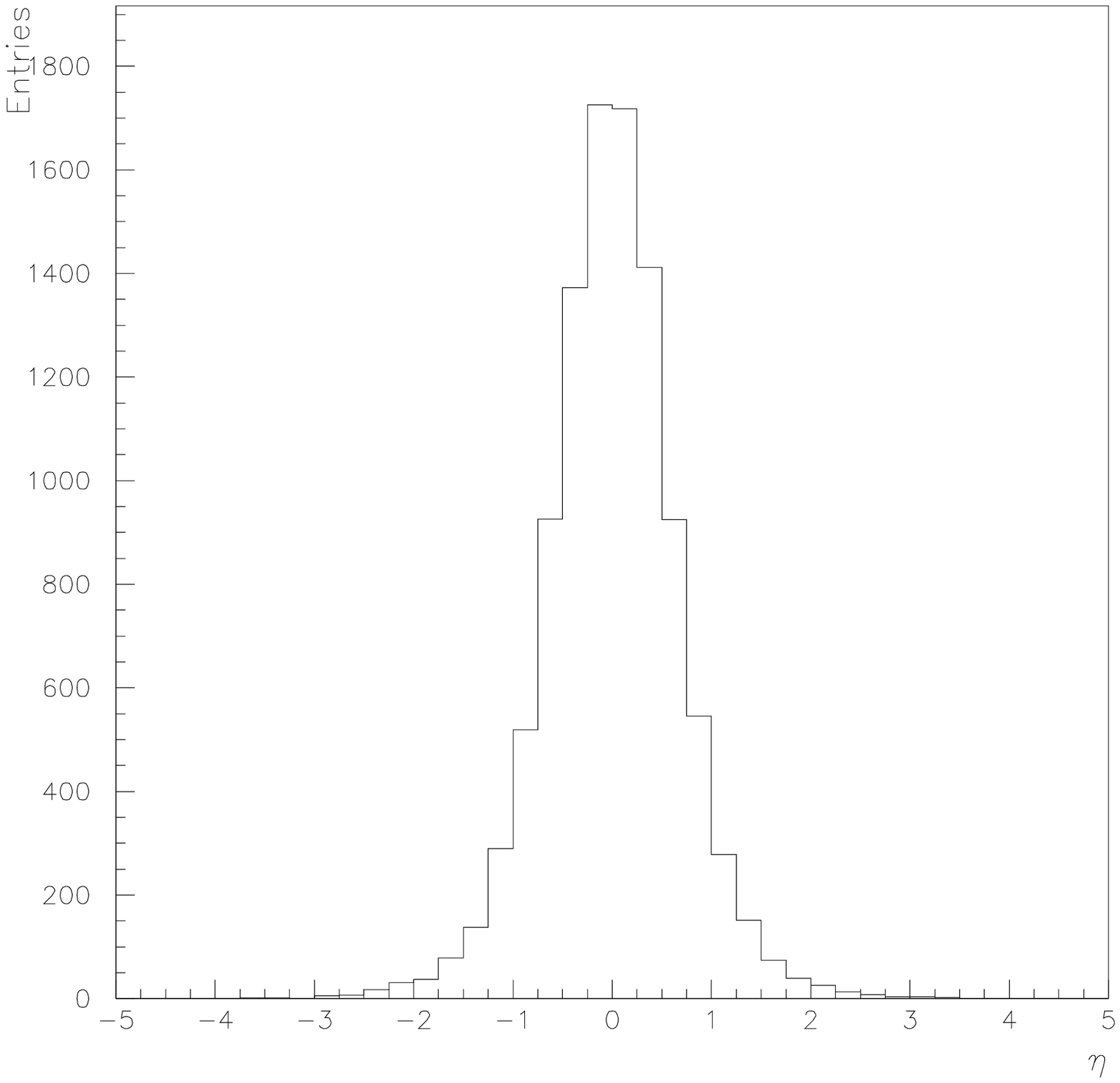}\includegraphics[scale=0.25]{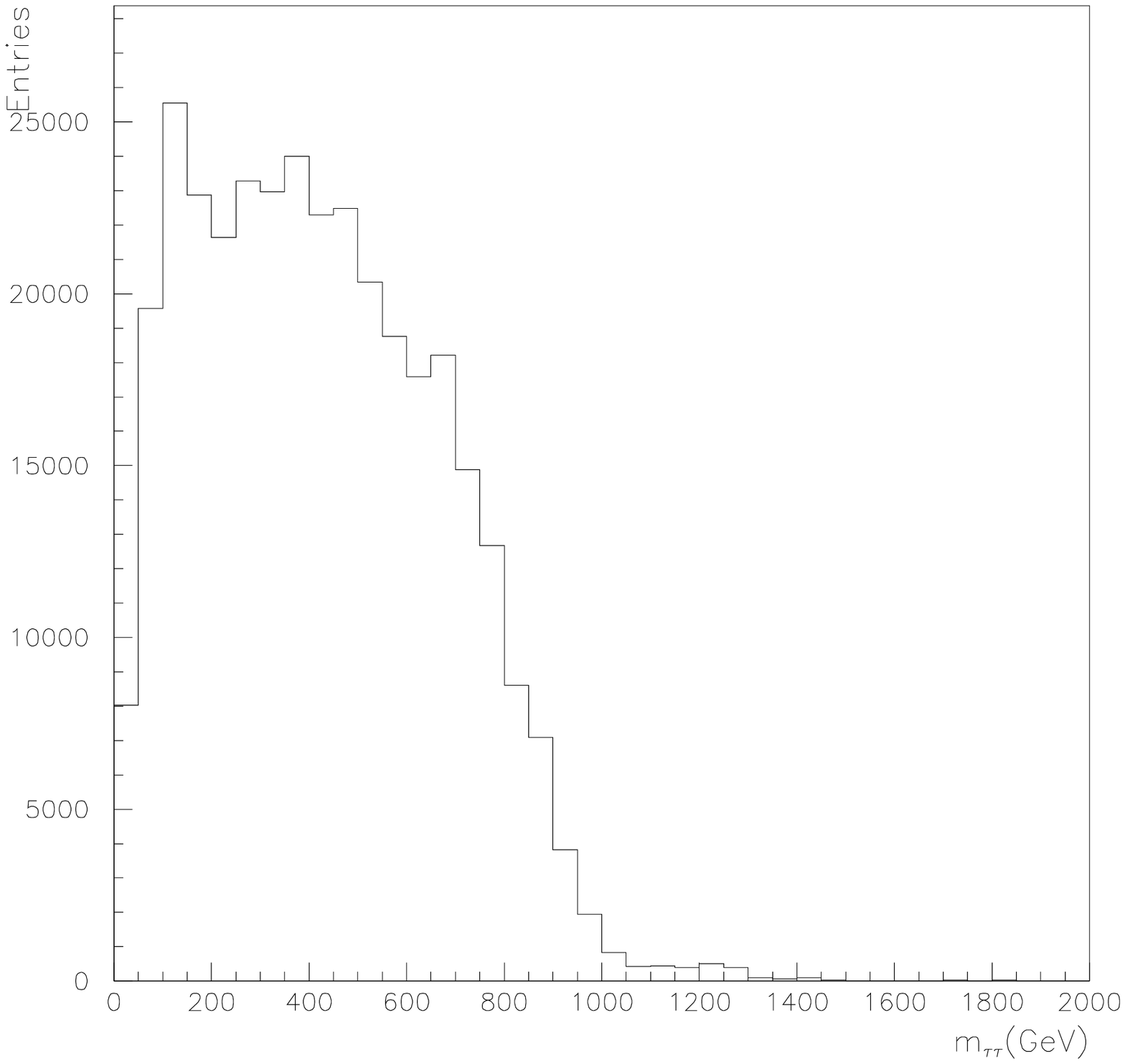}\caption{The same as Fig. \ref{fig:fig7}, but for the point $\beta$. \label{fig:fig8}}

\end{figure}

\begin{figure}
\includegraphics[scale=0.25]{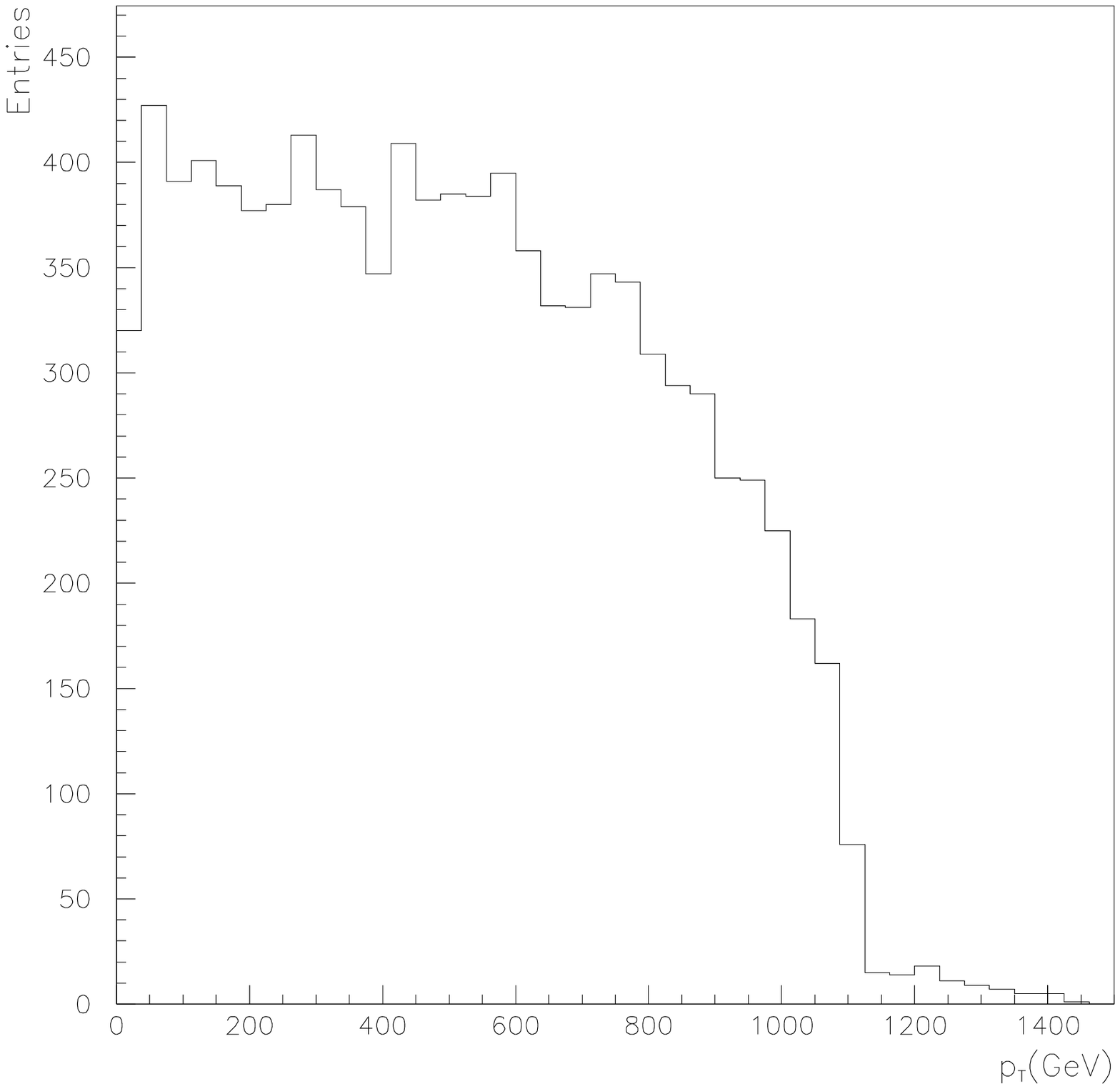}\includegraphics[scale=0.25]{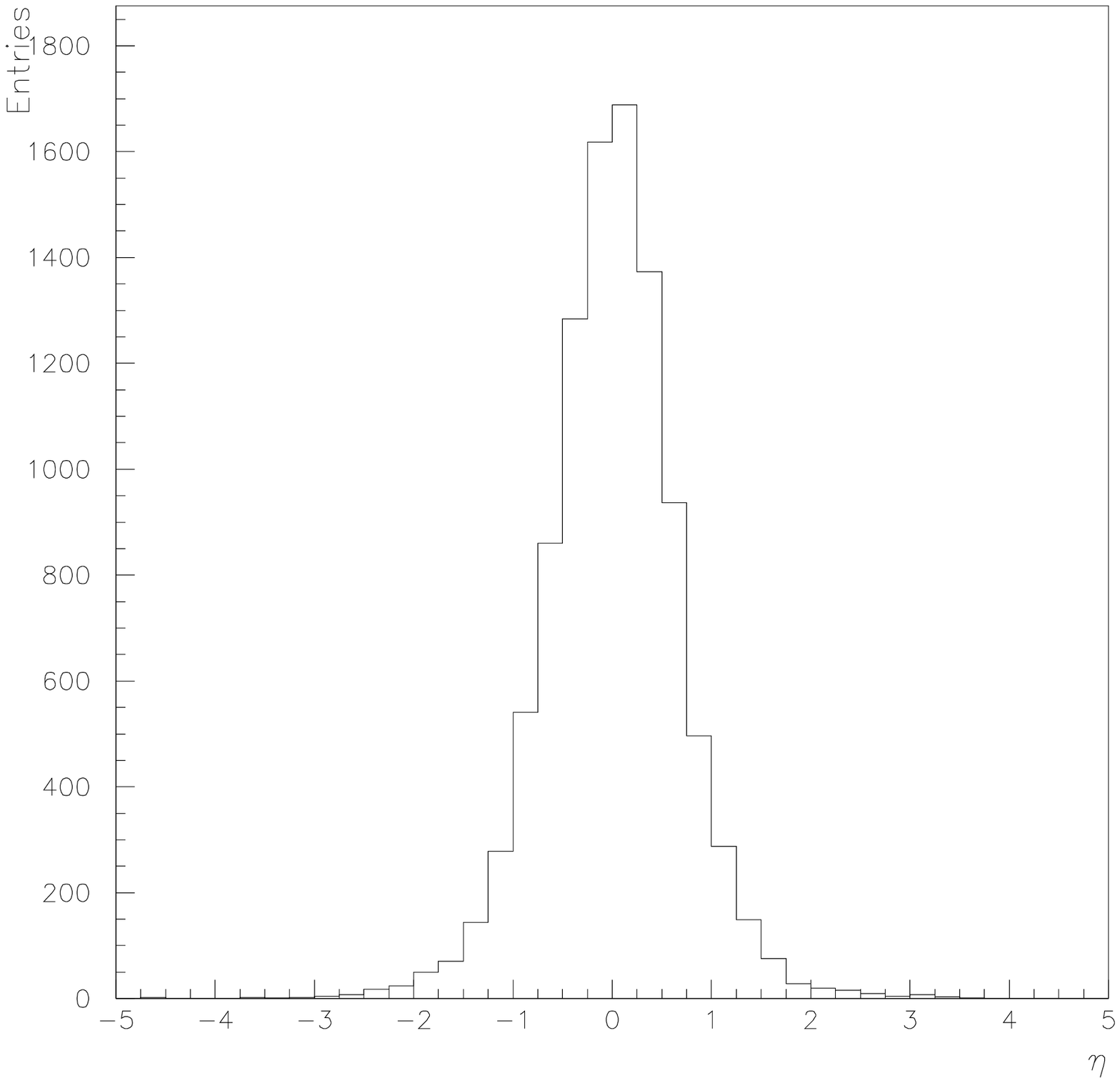}\includegraphics[scale=0.25]{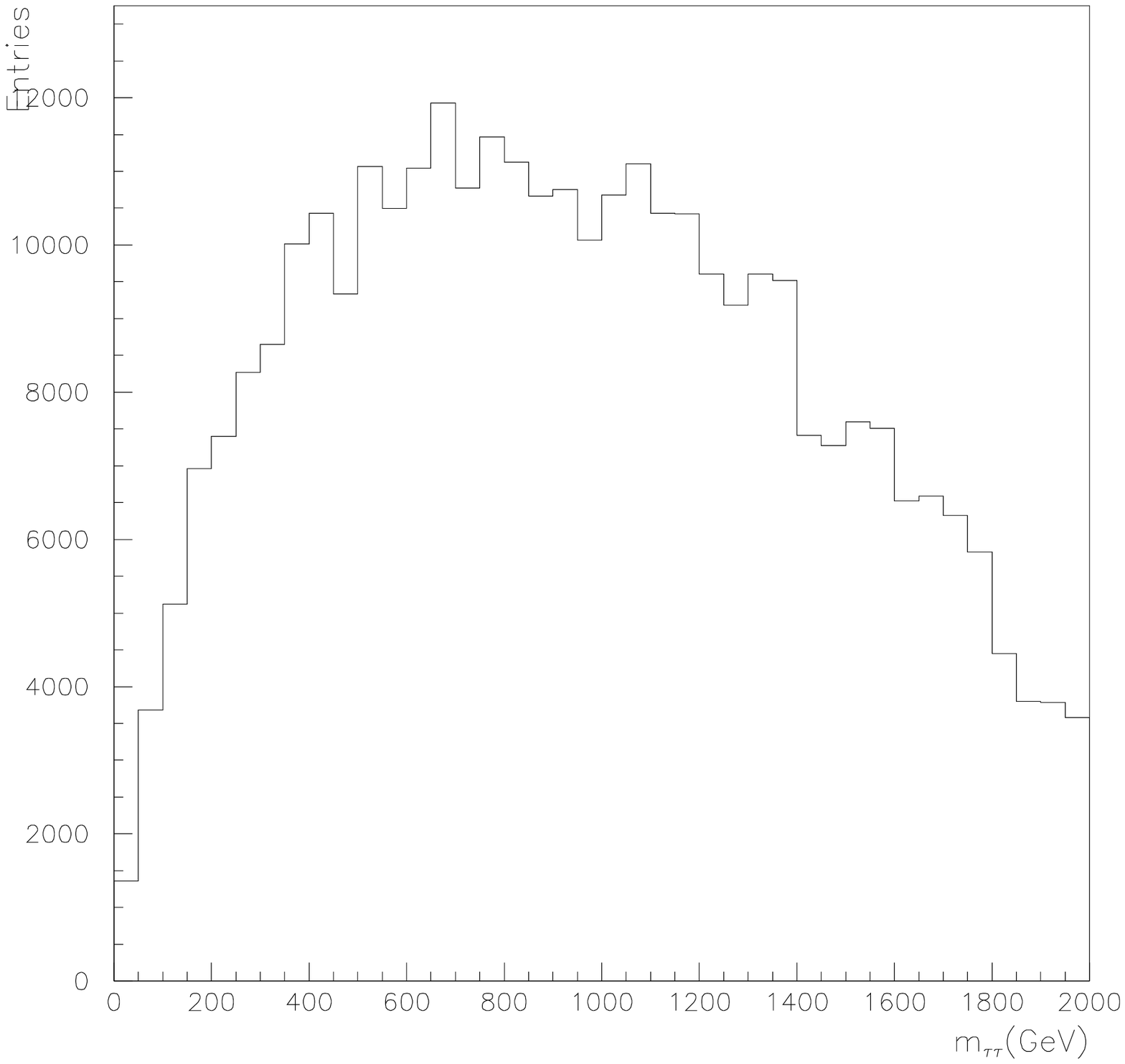}\caption{The same as Fig. \ref{fig:fig8}, but for the point $\gamma$. \label{fig:fig9}}

\end{figure}

\begin{figure}
\includegraphics[scale=0.25]{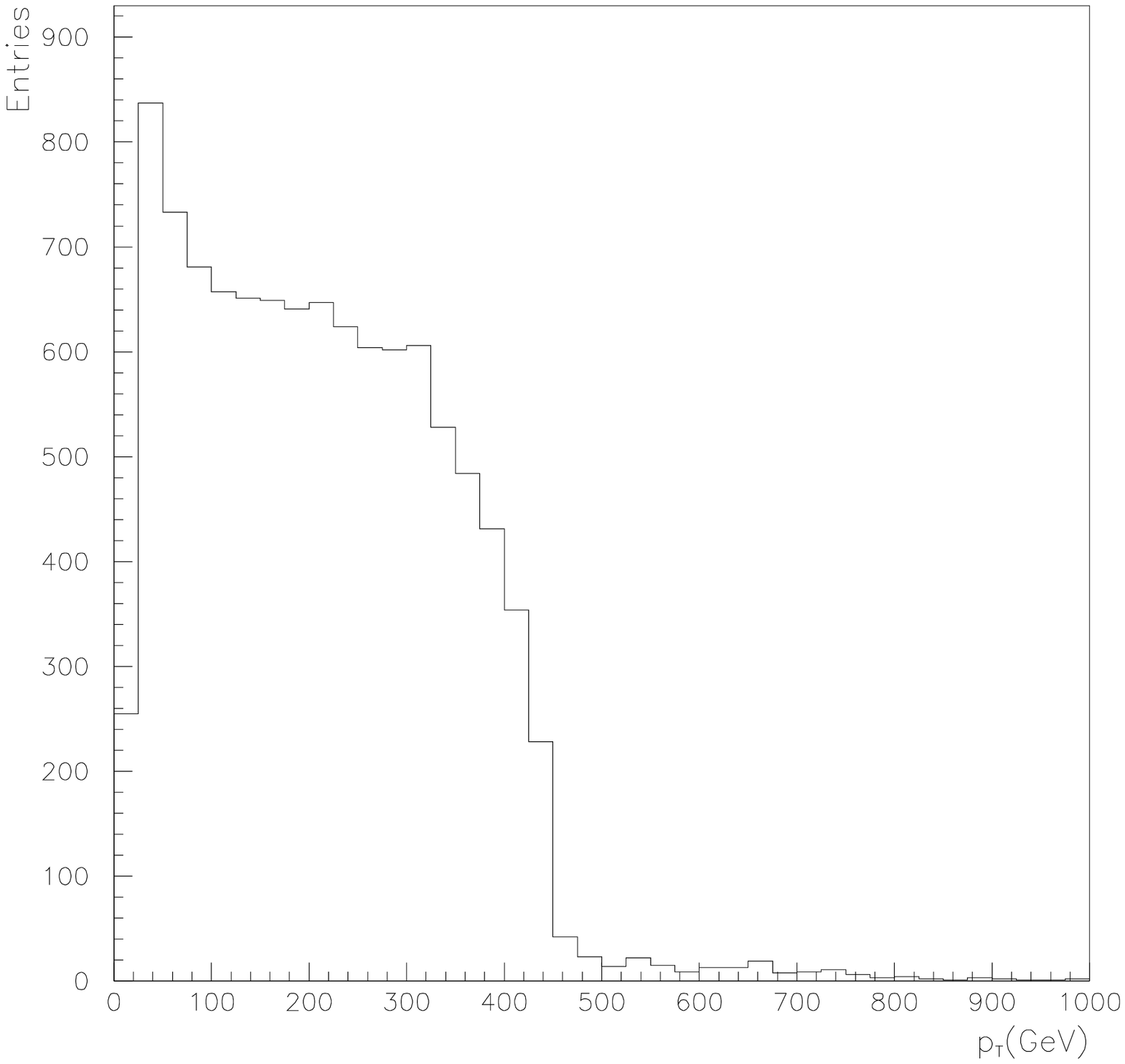}\includegraphics[scale=0.25]{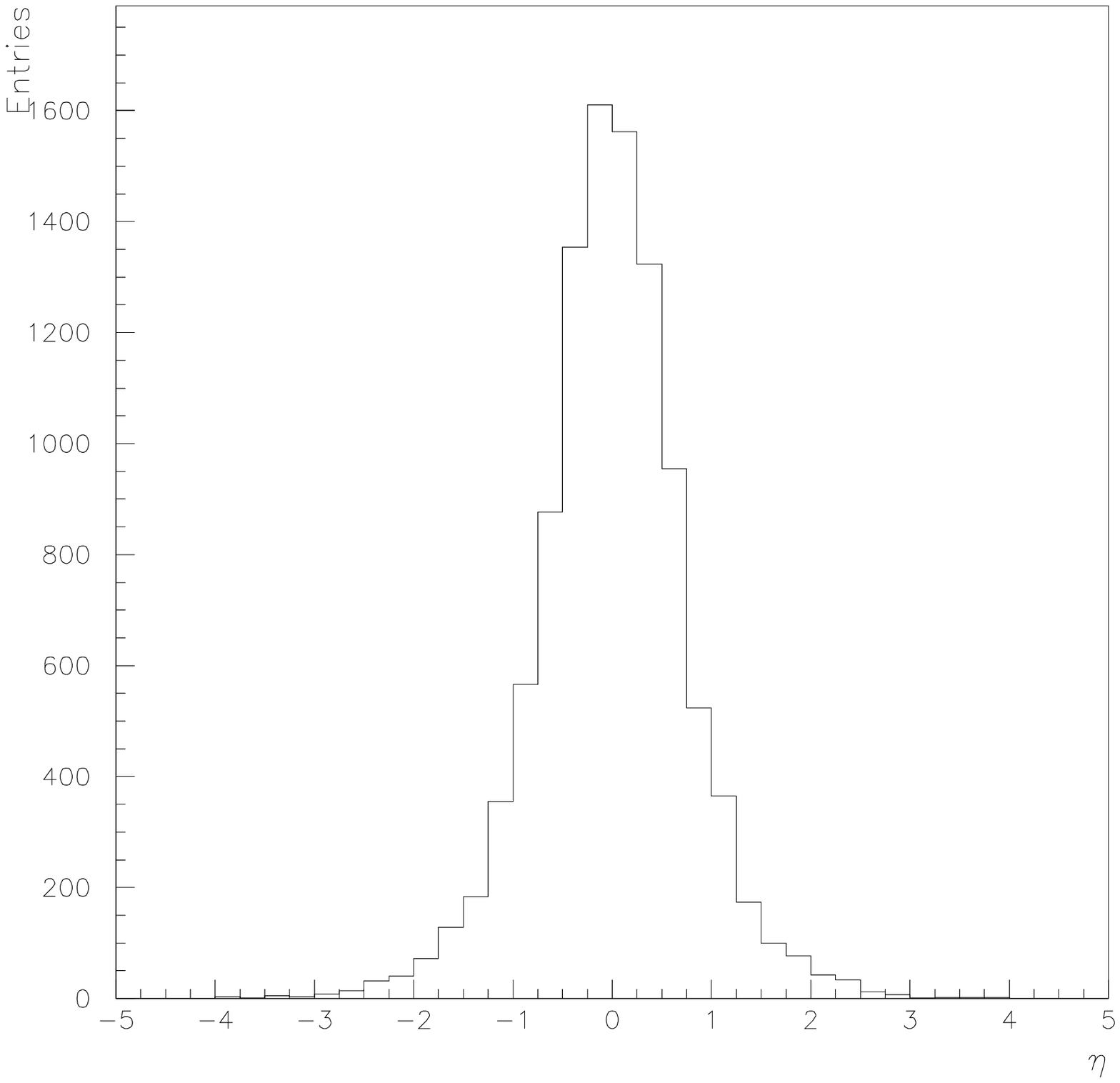}\includegraphics[scale=0.25]{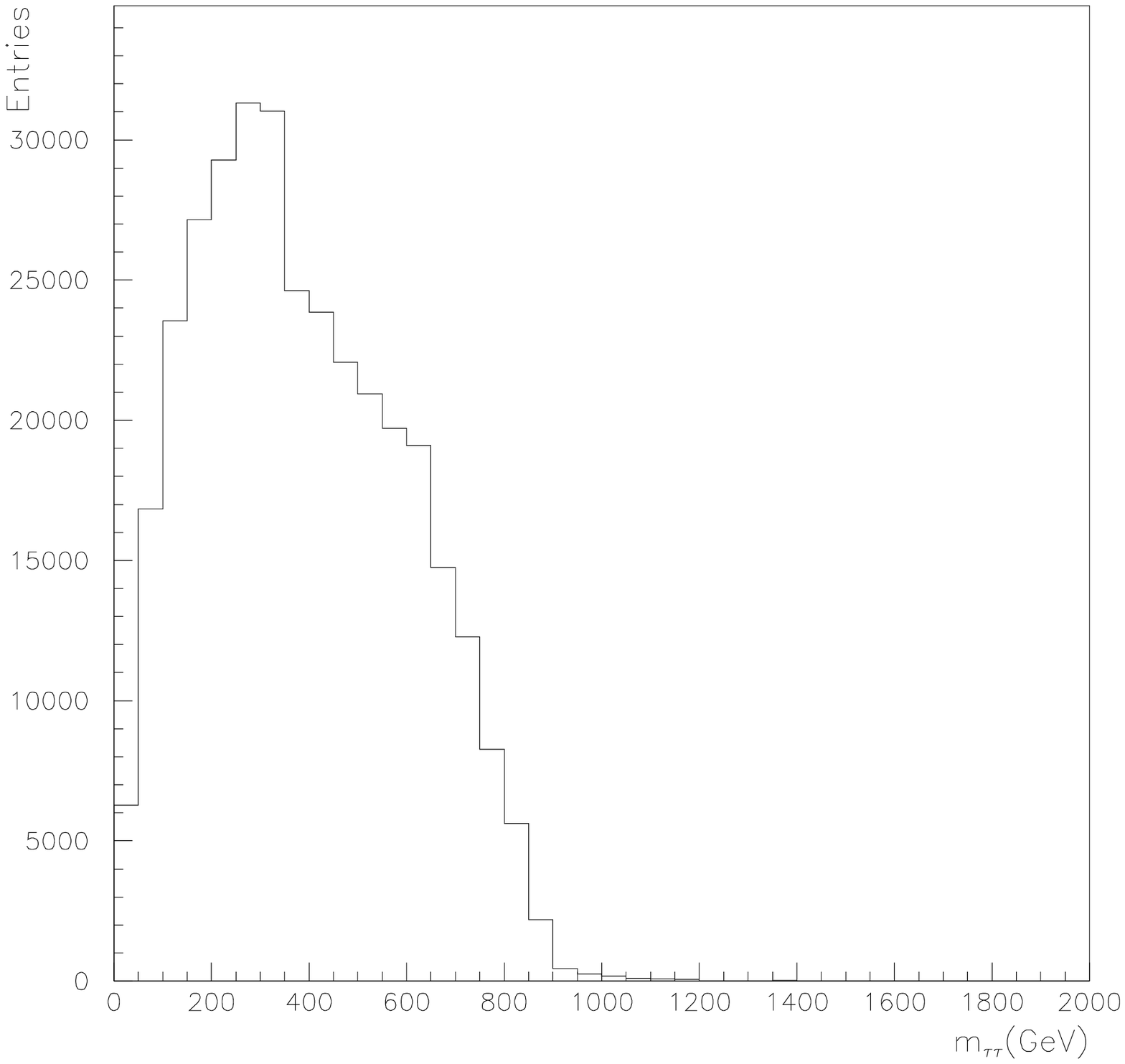}\caption{The same as Fig. \ref{fig:fig9}, but for the point $\delta$. \label{fig:fig10}}

\end{figure}

The distributions of the transverse momentum and rapidity of the $\tau$
lepton, and the invariant mass of two different sign $\tau$ leptons
for the points $\alpha$, $\beta$, $\gamma$ and $\delta$ are shown
in Figs. \ref{fig:fig7}-\ref{fig:fig10}. It is shown that $p_{T}$
distributions of the $\tau$-lepton have an upper edge around $650$
GeV, $500$ GeV, $1100$ GeV and $450$ GeV for the points $\alpha$,
$\beta$, $\gamma$ and $\delta$, respectively. The invariant mass
distributions of two $\tau$-leptons have peaks between 300 GeV and
700 GeV, depending on the points. For a contributing background process
$e^{+}e^{-}\to\tau^{+}\tau^{-}W^{-}W^{+}$, we find the $p_{T}$ distribution
of tau lepton decreasing smoothly in the range $p_{T}>50$ GeV. The
rapidity of tau lepton coming from this background shows a wide spectrum
($|\eta|<3$) different from the signal expected, and the tau leptons
originating from the $Z$ decays show an invariant mass spectrum around
$m_{Z}$.

Concerning the benchmark points, the tau sneutrinos are heavier than
the lightest chargino, the sneutrino can decay into a charged lepton
and a chargino $\tilde{\nu}_{\tau}^{(*)}\to\tau^{-(+)}\tilde{\chi}_{1}^{+(-)}$
, and chargino decays through lightest neutralino $\chi_{1}^{+(-)}\to\tilde{\chi}_{1}^{0}W^{-(+)}$
followed by the $W^{-(+)}$-boson decay to two jets or $l^{-(+)}+\not\! E_{T}$.
The branching ratio can be identified for these decays. Here, we consider
the hadronic decays of $W$-boson, then at the final state there will
be two opposite sign tau leptons, four jets and missing transverse
energy (MET). The tau polarization can be measured from the energy
distribution of the decay products of $\tau$. The degree of $\tau$
polarization can be used to determine the asymmetries and parameters
in supersymmetric processes \cite{Nojiri95,Nojiri96,Godbole05,Choi07,Dreiner10}.
Analyzing tau polarization will also give some information about the
sneutrino interactions through the process $\tilde{\nu}_{\tau}\to\tau^{-}\tilde{\chi}_{1}^{+}$.
Taking $\tau^{-}$($\tau^{+}$ has opposite polarization), we define
the polarization in terms of the difference of the number ($N$) of
left- and right-handed $\tau^{-}$ produced for the signal:

\begin{equation}
P_{\tau}=\frac{N_{L}-N_{R}}{N_{L}+N_{R}}=\frac{|C_{L}|^{2}-|C_{R}|^{2}m_{\tau}^{2}/2m_{W}^{2}\cos^{2}\beta}{|C_{L}|^{2}+|C_{R}|^{2}m_{\tau}^{2}/2m_{W}^{2}\cos^{2}\beta}\label{eq:7}\end{equation}

If we calculate the number of tau events as $N=\sigma\cdot BR\cdot\epsilon\cdot L$,
we find the degree of polarization for the signal. Here, we use the
interaction term from Eq. \ref{eq:3}, and find the tau polarization
in terms of the chiral couplings $C_{L,R}$. Since the right-handed
coupling is proportional to the mass ratio of $m_{\tau}/m_{W}$, which
is negligible, therefore the left-handed polarization will dominate
for the signal. We calculate $P_{\tau}=0.96$ for the point $\gamma$,
and $P_{\tau}=0.99$ for other points. Due to the different aspects
of the signal and background, an appropriate $p_{T}$ cut, pseudorapidity
cut and the invariant mass ($m_{\tau\tau}$) cut will be useful to
reduce the backgrounds.

\begin{figure}
\includegraphics[scale=0.6]{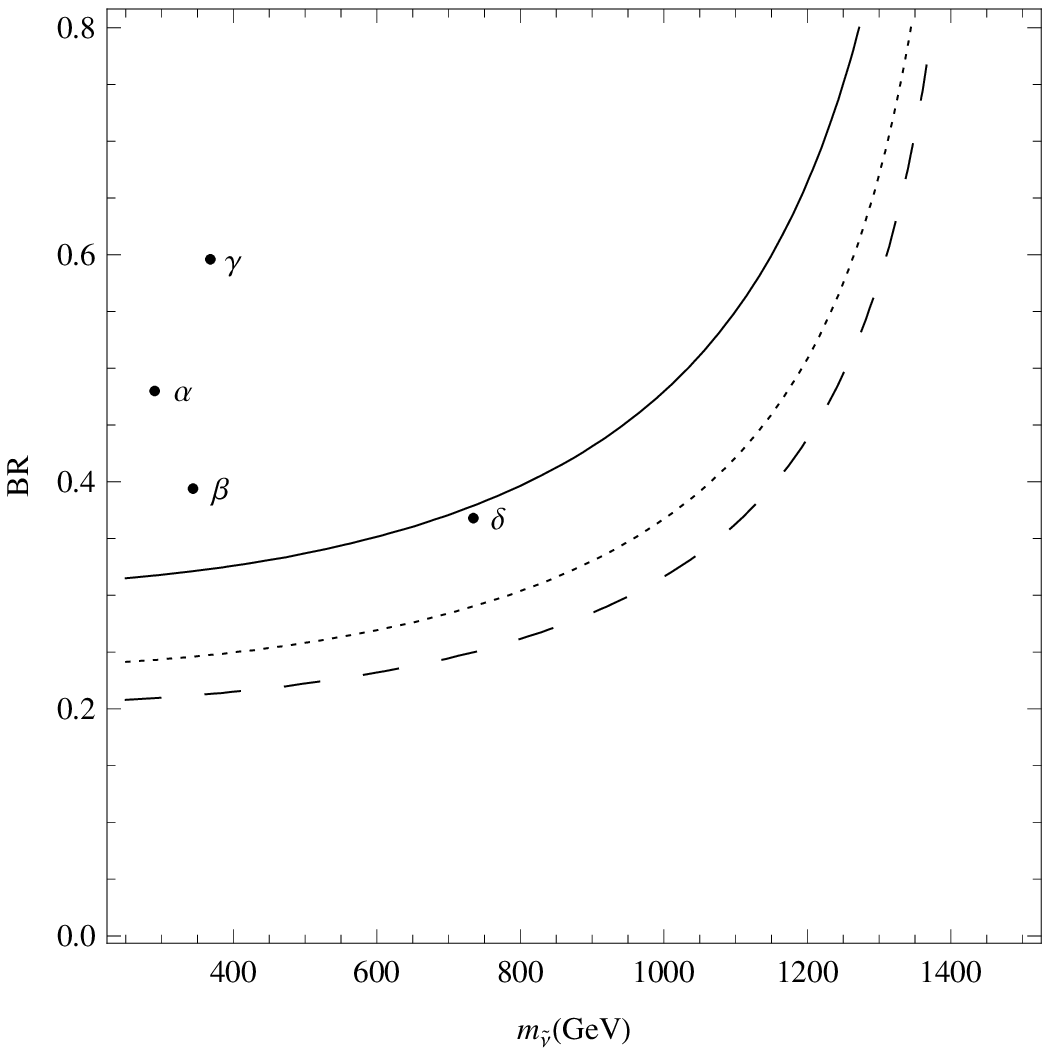} \includegraphics[scale=0.6]{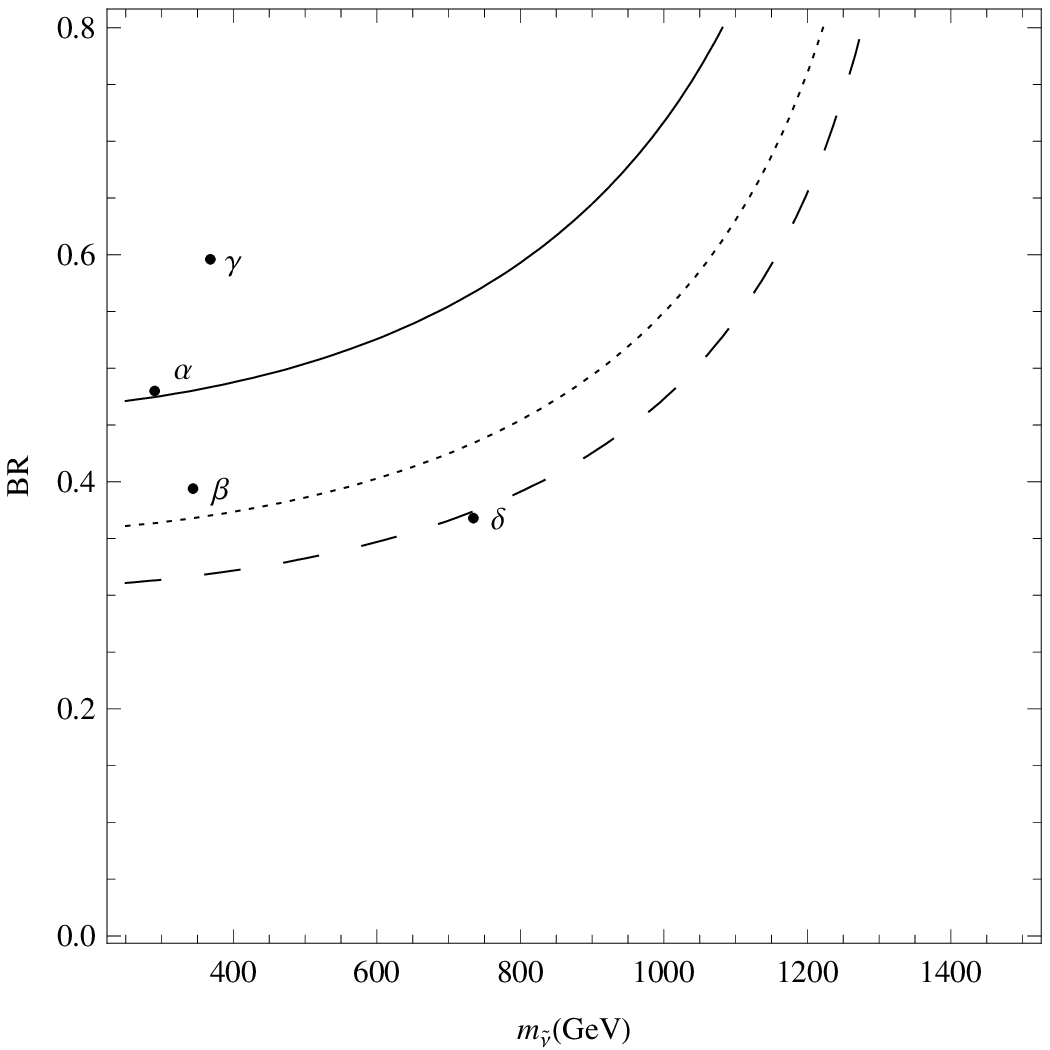}

\caption{Contour plot for branching ratio depending on the tau sneutrino mass
for the luminosities $L_{int}=1000$ fb$^{-1}$(left) and $L_{int}=200$
fb$^{-1}$(right) at the center of mass energy $\sqrt{s}=3000$ GeV.
The solid line corresponds to the unpolarized beams, dashed lines
show $e_{R}^{+}e_{L}^{-}$ case and dotted lines denotes $e_{L}^{+}e_{R}^{-}$.\label{fig:fig11}}

\end{figure}

In Fig. \ref{fig:fig11}, we plot the contour lines in the plane of
the branching ratio - tau sneutrino mass for the luminosities $L_{int}=1000$
fb$^{-1}$(left) and $L_{int}=200$ fb$^{-1}$(right) at the center
of mass energy $\sqrt{s}=3000$ GeV. The solid line corresponds to
the unpolarised positron and electron beams, dashed lines show $e_{R}^{+}e_{L}^{-}$
case and dotted line denotes $e_{L}^{+}e_{R}^{-}$. The beam polarisation
is helpful in the study of SUSY processes to improve the $S/\sqrt{B}$
(analyser). We consider here three options: unpolarised, LR and RL
polarised positron/electron beams. For an integrated luminosity of
$1$ ab$^{-1}$, it is possible to cover four points even at the unpolarised
case. In Table \ref{tab:4}, we present the integrated luminosity
required to get $3\sigma$ signal significance at $\sqrt{s}$=3 TeV. 

\begin{table}
\caption{The luminosity (in fb$^{-1}$) required to obtain a $3\sigma$ signal
significance at 3 TeV. \label{tab:4}}

\begin{tabular}{|c|c|c|c|c|}
\hline 
Beams\textbackslash{}points & $\alpha$ & $\beta$ & $\gamma$ & $\delta$\tabularnewline
\hline 
$e^{+}e^{-}$ & 191 & 443 & 86 & 1118\tabularnewline
\hline 
$e_{L}^{+}e_{R}^{-}$ & 65 & 152 & 29 & 385\tabularnewline
\hline 
$e_{R}^{+}e_{L}^{-}$ & 36 & 84 & 16 & 212\tabularnewline
\hline
\end{tabular}
\end{table}

In conclusion, tau sneutrino pair production could give a valuable
information about the sneutrino interactions. The LHC has the potential
to measure the SUSY mass spectrum and a clue on the underlying scenarios
with the exploitation of full high luminosity. A more precise determination
of specific processes can be performed within the underlying model
at future lepton colliders operating with polarized beams.
\begin{acknowledgments}
The numerical calculations reported in this paper were performed at
TUBITAK ULAKBIM, High Performance and Grid Computing Center (TR-Grid
e-Infrastructure). This work is partially supported by Turkish Atomic
Energy Authority (TAEK). O.C\textquoteright{}s work is partially supported
by State Planning Organization (DPT) under the grant No. DPT2006K-120470.\end{acknowledgments}

\end{document}